\documentclass[aps,prb,twocolumn,showpacs,byrevtex]{revtex4}
\usepackage{graphicx}

\begin{document}
\title{STM characterization of the Si-P heterodimer}
\author{N.~J.~Curson}
\email{neil.curson@unsw.edu.au} \affiliation{Centre for Quantum
Computer Technology, School of Physics, University of New South
Wales, Sydney 2052, Australia}
\author{S.~R.~Schofield}
\affiliation{Centre for Quantum Computer Technology, School of
Physics, University of New South Wales, Sydney 2052, Australia}
\author{M.~Y.~Simmons}
\affiliation{Centre for Quantum Computer Technology, School of
Physics, University of New South Wales, Sydney 2052, Australia}
\author{L.~Oberbeck}
\affiliation{Centre for Quantum Computer Technology, School of
Physics, University of New South Wales, Sydney 2052, Australia}
\author{J.~L.~O'Brien}
\altaffiliation{Present address: Centre for Quantum Computer
Technology, Department of Physics, University of Queensland,
Brisbane 4072, Australia}
\author{R.~G.~Clark}
\affiliation{Centre for Quantum Computer Technology, School of
Physics, University of New South Wales, Sydney 2052, Australia}

\begin{abstract}
We use scanning tunneling microscopy (STM) and Auger electron
spectroscopy to study the behavior of adsorbed phosphine
(PH$_{3}$) on Si(001), as a function of annealing temperature,
paying particular attention to the formation of the Si-P
heterodimer. Dosing the Si(001) surface with ${\sim}$0.002
Langmuirs of PH$_{3}$ results in the adsorption of PH$_{x}$
(x=2,3) onto the surface and some etching of Si to form individual
Si ad-dimers. Annealing to 350$^{\circ}$C results in the
incorporation of P into the surface layer to form Si-P
heterodimers and the formation of short 1-dimensional Si dimer
chains and monohydrides. In filled state STM images, isolated Si-P
heterodimers appear as zig-zag features on the surface due to the
static dimer buckling induced by the heterodimer. In the presence
of a moderate coverage of monohydrides this static buckling is
lifted, rending the Si-P heterodimers invisible in filled state
images. However, we find that we can image the heterodimer at all
H coverages using empty state imaging. The ability to identify
single P atoms incorporated into Si(001) will be invaluable in the
development of nanoscale electronic devices based on controlled
atomic-scale doping of Si.
\end{abstract}

\maketitle

\section{Introduction}

Currently there are several proposals to use phosphine gas
(PH$_{3}$) to create atomic-scale devices by positioning P atoms
on a surface using
H-lithography.~\cite{tu-ijcta-00-553,ob-prb-01-161401,sc-prl-03-136104}
One such idea~\cite{ob-prb-01-161401} involves the fabrication of
a 2-dimensional array of P atoms in Si for the realization of a
silicon based quantum computer.~\cite{ka-na-98-133} A requirement
of each of these schemes is that the P atoms, relinquished from
the PH$_{3}$ molecules, are incorporated into substitutional
lattice sites. Since PH$_{3}$ is used for doping Si in the
semiconductor industry, the interaction of PH$_3$ with the Si(001)
surface has been studied for over two decades using a wide variety
of surface sensitive techniques (e.g.,
Ref.~\onlinecite{yu-jvsta-84-446}). It is well established that
the adsorption of PH$_3$ on the Si(001) surface is partially
dissociative, resulting in the adsorption of both PH$_3$ and
PH$_2.$~\cite{co-jvsta-94-2995,yo-jap-95-4988,ha-ass-96-25,sh-jpc-96-4961,li-ss-99-7,sc-prl-03-136104}
Moreover, it is known that annealing PH$_3$ dosed surfaces causes
the complete dissociation of the adsorbed PH$_3$ and PH$_2$
molecules (estimates of the temperature at which this occurs vary
within the range 300$^\circ$C -
425$^\circ$C~\cite{sh-jpc-96-4961,ha-ass-96-25,li-prb-00-2799,co-jvsta-94-2995})
resulting in the incorporation of P into the surface to form Si-P
heterodimers.~\cite{wa-prb-94-4534,ha-ass-96-25,lo-apl-02-3197,sc-prl-03-136104}
The majority of the investigations mentioned above have studied
the interaction of phosphine with the Si(001) surface at high
surface coverages. However, at low coverages there are only two
papers reporting detailed atomic-resolution scanning tunneling
microscopy (STM) investigations of the Si(001):PH$_3$ system, the
first by Wang~\textit{et al.}~\cite{wa-jpc-94-5966} and the second
by Kipp~\textit{et al.}~\cite{ki-prb-95-5843} While the
experimental results presented in these two papers are in good
agreement, their interpretations are contradictory.

The experimental results from Wang~\textit{et
al.}~\cite{wa-jpc-94-5966} and Kipp~\textit{et
al.}~\cite{ki-prb-95-5843} showed that after the Si(001) surface
was exposed to a low dose of PH$_{3}$, two types of bright
features appeared on the surface. Both features were centered on
top of the Si dimer rows, with one feature larger than the other.
Wang~\textit{et al.}~\cite{wa-jpc-94-5966} also observed an
increase in the number of dimer vacancies on the surface after
PH$_{3}$ dosing. In both studies, after the dosed surfaces were
annealed (to 230$^\circ$C, Ref~\onlinecite{wa-jpc-94-5966}; and to
350$^\circ$C,~Ref~\onlinecite{ki-prb-95-5843}), the two types of
bright feature disappeared from the surface and were replaced by
1-dimensional (1-D) chains running perpendicular to the underlying
Si dimer rows and a few small 2-dimensional (2-D) islands. The
above data was interpreted by Wang~\textit{et
al.}~\cite{wa-jpc-94-5966} in the following way: the larger bright
feature was a single Si atom (a monomer) ejected from the surface
and the smaller bright feature was a PH$_{3}$ molecule adsorbed on
top of a Si dimer. Upon heating the surface to 230$^{\circ}$C the
Si monomers combined to form 1-D rows and small 2-D islands. This
annealing also induced complete PH$_{3}$ dissociation, with the P
atoms being incorporated into the surface and the H adsorbed onto
the surface to form monohydride dimers. Kipp~\textit{et
al.}~\cite{ki-prb-95-5843} presented a different interpretation of
the above data. They concluded that the large bright features were
PH$_{n}$ (n=1-3) molecules adsorbed at defects and the small
bright features were P-P dimers, resulting from the pairing and
subsequent dissociation of two PH$_{3}$ molecules (with the
dissociated hydrogen atoms also bonding to the surface). Upon
heating the surface to 350$^{\circ}$C Kipp~\textit{et
al.}~\cite{ki-prb-95-5843} concluded that the P-P dimers coalesce
to form small chains and islands. They suggested that the
formation of large islands was hindered by the presence of
adsorbed H on the surface.

In contrast to the results of Wang~\textit{et al.} and
Kipp~\textit{et al.}, separate Fourier transform infrared
spectroscopy studies~\cite{sh-jpc-96-4961} and a combined electron
energy loss spectroscopy and thermal programmed desorption
study~\cite{co-jvsta-94-2995} have indicated that molecular
PH$_{3}$ adsorption is accompanied by a partial dissociation of
PH$_{3}$ to form adsorbed PH$_{2}$ and H, during phosphine
adsorption at room temperature. Therefore, further study of this
system is required to consolidate the different interpretations.
We have used high resolution STM and Auger electron spectroscopy
to study the evolution of a PH$_{3}$ dosed Si(001) surface, as a
function of temperature and coverage. Particular emphasis was
placed on the process of P incorporation into the Si(001) surface
and the formation of the Si-P heterodimer. To this end we have
also performed a study of surface segregated P from a Si
encapsulated $\delta$-doped layer. Pivotal to these experiments is
the ability to resolve the presence of a single P atom in the
Si(001) surface using STM. We show that the isolated Si-P
heterodimer appears as a zig-zag feature in filled state STM
images, but that the heterodimer is not visible if there is a
moderate coverage of monohydrides on the surface. However, empty
state imaging can be used to identify the heterodimers at all
coverages.

\section{Experimental}

Experiments were performed in ultra-high vacuum (base pressure
$<5\times10^{-11}$~mbar) using an Omicron variable temperature STM
and electrochemically etched tungsten tips. The n-type Si(001)
sample had a resistivity of ${\sim}$ 1 ${\Omega}$cm and was
prepared in vacuum by out-gassing overnight at 575$^{\circ}$C
using a resistive heater element, followed by flashing to
1175$^{\circ}$C by passing a direct current through the sample.
After flashing, the samples were cooled slowly
(${\sim}$3$^{\circ}$C/s) from 900$^{\circ}$C to room temperature.
For all anneal steps the sample temperature was maintained at the
stated anneal temperature for 10 seconds unless stated otherwise.
The PH$_{3}$ dosing was performed by opening a precision leak
valve, which faced the sample, between the UHV chamber and a
PH$_{3}$ micro-dosing system. All doses were performed by
back-filling the chamber with a 1 ${\times}$ 10$^{-9}$ Torr
pressure of PH$_{3}$ and controlling the dose in Langmuirs (1 L =
10$^{6}$ ${\times}$ pressure (torr) ${\times}$ dose (seconds)) via
the dose time. The ionization gauge was not in line-of-sight of
the sample. All STM images were taken at room temperature and are
filled state images (negative sample bias) unless stated
otherwise. Dual bias STM images, where filled and empty state
images are obtained simultaneously, were taken with a negative
(positive) sample bias while scanning the tip in forward (reverse)
direction.

\section{Results and discussion}
\subsection{Identification of surface features for room temperature adsorption of PH$_{3}$ on Si(001)}

\begin{figure}
\begin{center}
\includegraphics[width=8cm]{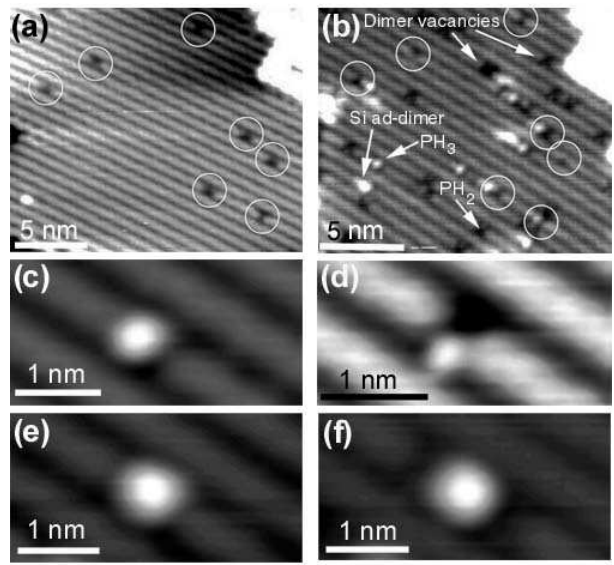}
  \caption{\label{doseFig}STM images of the same area of a Si(001) surface (a) before and
(b) after dosing with 0.002 L of PH$_{3}$ at room temperature.
Imaging conditions were -1.6 V sample bias and 0.2 nA tunneling
current. (c) and (d) show PH$_{3}$ and PH$_{2}$ + H, adsorbed on
Si(001), respectively. Imaging conditions were -1.8 V sample bias
and 0.1 nA tunneling current. High resolution STM images of a Si
ad-dimer are shown for (e) filled state (-1.6 V, 0.1 nA) and (f)
empty state (+1.2 V and 0.1 nA) imaging.}
\end{center}
\end{figure}

Figure~\ref{doseFig} shows STM images of the Si(001) surface after
exposure to low coverages of phosphine (PH$_{3}$) gas, at room
temperature. The clean Si(001) surface consists of rows of Si
dimers which are seen as the parallel lines running diagonally
across the images.
Figure~\ref{doseFig}(a) and (b) show a 21 nm ${\times}$ 17 nm area
of a Si(001) surface before and after dosing with 0.002 L of
PH$_{3}$, respectively. The circled features in (a) are single
dimer vacancy defects which are inherent to the clean surface and
which we identify to distinguish them from species resulting from
PH$_{3}$ dosing. After PH$_{3}$ dosing [Fig.~\ref{doseFig}(b)] we
find that additional features have appeared on the surface, which
are: (i) adsorbed PH$_{3}$, (ii) adsorbed PH$_{2}$ + H, (iii)
dimer vacancies (in addition to those present before dosing) and
(iv) Si ad-dimers. Figure~\ref{doseFig}(c) shows an image of a
PH$_{3}$ molecule which has a height/diameter of 0.9~\AA/11~\AA~in
filled state images and is adsorbed upon the center of a dimer
row, as previously identified by Wang, Shan and
coworkers.~\cite{wa-jpc-94-5966,sh-jpc-96-4961} It can be seen in
Fig.~\ref{doseFig}(c) that this molecule is adsorbed adjacent to a
defect, which we find is commonly the case at low coverage. We
believe that defects stabilize the adsorption of the PH$_{3}$
molecules centered over Si dimers. However we cannot rule out the
possibility that the feature is a PH$_{2}$ radical centered on the
dimer row, as suggested by Lin~\textit{et al.}~\cite{li-ss-99-7}
Figure~\ref{doseFig}(d) shows a PH$_{3}$ molecule after it has
dissociated into PH$_{2}$ + H on a single Si dimer, as
described.~\cite{sc-prl-03-136104} The observation of both
adsorbed PH$_{3}$ in its molecular form, and as the dissociation
products PH$_{2}$ + H is in agreement with
FTIR~\cite{sh-jpc-96-4961} and photoemission~\cite{li-ss-99-7}
results.

The third type of features in Figure~\ref{doseFig}(b) are single
and multiple dimer vacancies, resulting directly from the removal
of Si from the surface layer during PH$_{3}$ dosing. While dimer
vacancies are difficult to distinguish from monohydride dimers
(formed when each Si atom of a dimer is terminated with an H atom)
the formation of monohydride dimers is not expected at room
temperature since single H atoms have insufficient mobility to
diffuse and pair-up into monohydrides.~\cite{bo-jpcm-00-7655} We
assign the fourth feature to be a Si ad-dimer, centered on the
dimer rows with a height/diameter of 1.3~\AA/13~\AA~in the filled
state image and a height/diameter of 1.9~\AA~/13~\AA~in the empty
state, in agreement with the observation of Si ad-dimers by
Swartzentruber~\cite{sw-ss-97-195} and shown in
Fig.~\ref{doseFig}(e) and (f) respectively. The filled state
height of the feature (1.3~\AA)~is of the same order as the
distance between [001] planes of the Si crystal (1.36~\AA)~as
expected for a Si ad-dimer. We discount the suggestion of
Wang~\textit{et al.}~\cite{wa-jpc-94-5966} that the feature is a
Si monomer, since Si monomers are not observed in filled state
images.~\cite{sw-jcg-98-1} We also rule out the fourth feature
being either adsorbed P or PH species because Fourier transform
infrared spectroscopy~\cite{sh-jpc-96-4961} and
photoemission~\cite{li-ss-99-7} results show that these species
are not formed after room temperature PH$_{3}$ adsorption on
Si(001).

\subsection{P identification using STM and AES}

\begin{figure}
\begin{center}
\includegraphics[width=8cm]{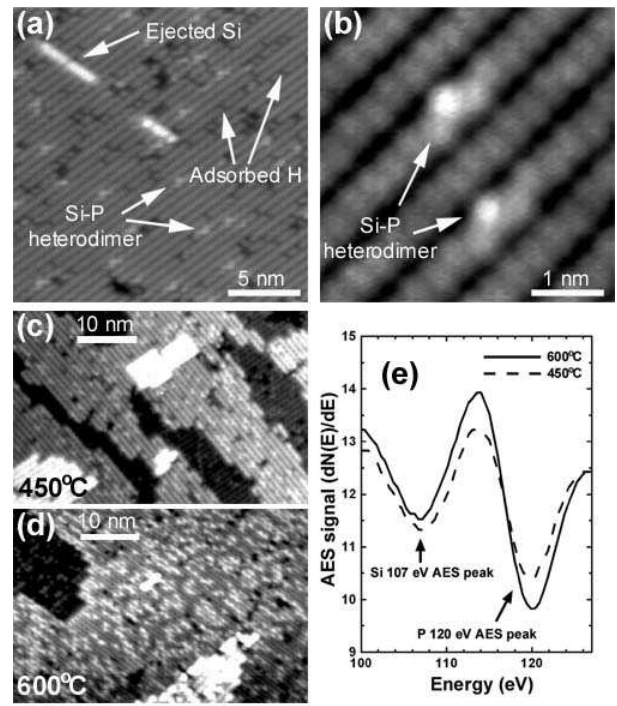}
  \caption{\label{350A}(a), (b) STM images of a Si(001) surface after dosing with 0.002 L of PH$_{3}$ and annealing to
  350$^{\circ}$C at low and high resolution respectively. (c),(d) STM images of a Si(001) surface after
saturation dosing with PH$_{3}$, annealing to 600$^{\circ}$C,
overgrowing with 5 ML of Si at room temperature and
  then annealing to (c) 450$^{\circ}$C and (d) 600$^{\circ}$C.
  (e) Auger electron spectra obtained from the surfaces
  shown in (c) and (d).}
\end{center}
\end{figure}

\begin{figure}
\begin{center}
\includegraphics[width=8cm]{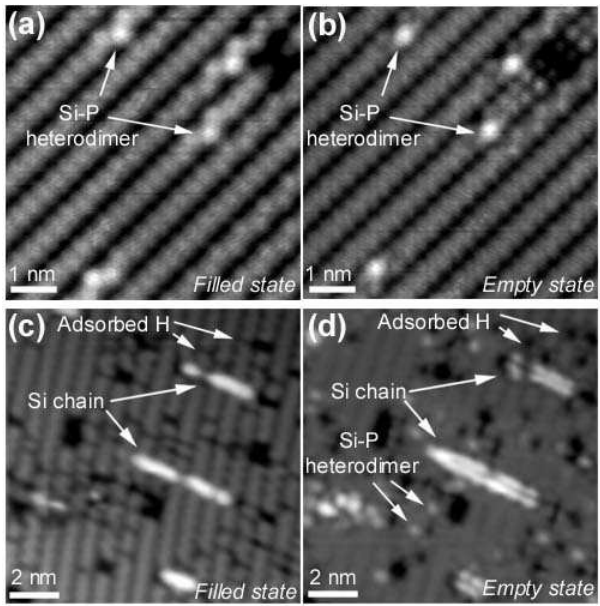}
  \caption{\label{dualbias}(a) Filled state and (b) empty state STM images of a Si(001)
  surface after dosing with 0.002 L of PH$_{3}$ and annealing to 500$^{\circ}$C. (c) Filled state and (d) empty state STM images of a Si(001)
  surface after dosing with 0.013 L of PH$_{3}$ and annealing to 350$^{\circ}$C.
  Tunneling conditions were (a) and (c) -1.6 V and 0.13 nA, (b) and (d) +1.2 V and 0.13 nA.}
\end{center}
\end{figure}

We now describe the incorporation of P atoms into the surface, to
form Si-P heterodimers, upon annealing. Fig.~\ref{350A}(a) shows a
Si(001) surface that has been exposed to 0.002 L of PH$_{3}$ and
annealed to 350$^{\circ}$C. Three distinct types of feature are
present on the surface, Si-P heterodimers, ejected Si dimer chains
and monohydride dimers. The ejected Si chains appear as short
bright lines running perpendicular to the dimer rows. These
features have the same height above the surface (1.36~\AA) as a
monolayer Si step~\cite{dab-ch3} and have a characteristic minima
running lengthways along their center in empty state images (see
Fig.~\ref{dualbias}(d)), indicating that they are indeed Si dimer
chains as suggested by Wang~\textit{et al.}~\cite{wa-jpc-94-5966}
and not P dimer chains as proposed by Kipp~\textit{et
al.}~\cite{ki-prb-95-5843} The monohydride dimers (H-Si-Si-H)
appear as dark features, similar in appearance to missing dimer
defects.
Fig.~\ref{350A}(b) shows a high resolution image of two Si-P
heterodimers, which appear as zig-zags along the dimer row, with
the highest part ${\sim}$ 0.3~\AA~above the surface
dimers.~\cite{lo-apl-02-3197} This zig-zag appearance indicates
there is static buckling of the surface dimers on either side of
the Si-P heterodimer. Figure~\ref{350A}(b) shows that this
buckling decays in amplitude as a function of distance from the
Si-P heterodimer and extends 2-3 dimers away from it.

To confirm the interpretation of Fig.~\ref{350A}(a) presented
above, we compare the number of Si-P heterodimers in STM images
with the fraction of P at the surface measured by Auger electron
spectroscopy (AES). In order to control the number of Si-P
heterodimers at the surface without the complication of ejected Si
dimer chains and adsorbed H (Fig.~\ref{350A}(a)), we have created
a buried P $\delta$-doped layer and then diffused varying amounts
of the P atoms to the surface using a range of anneal
temperatures. The $\delta$-layer was produced by saturation dosing
a Si(001) surface with PH$_{3}$, annealing to 600$^{\circ}$C, and
encapsulating with 5 monolayers of Si at room
temperature.~\cite{lo-prb-preparation} Upon the annealing of this
sample the buried P atoms diffuse to the surface at a rate
determined by the anneal temperature. Figures~\ref{350A}(c)
and~\ref{350A}(d) show STM images obtained after annealing the P
$\delta$-doped sample for 60 seconds at 450$^{\circ}$C and
600$^{\circ}$C respectively. Figure ~\ref{350A}(e) shows the 120
eV P AES peak obtained from the two surfaces. There is a direct
correlation between the number of zig-zag features in the two STM
images and the intensity of the 120 eV P Auger peak, confirming
that the zig-zag feature contain P atoms. From the zig-zag
appearance we conclude that each of these features contain only
one P atom since P-P dimers are known to have have mirror symmetry
about a line running along their center~\cite{wa-prb-94-4534} and
a feature consisting of neighboring incorporated P atoms
$\emph{along}$ would show neighboring atoms with similar
intensities.

While the results presented here are the first to characterize the
isolated Si-P heterodimer, the observation that P atoms
incorporate into the Si(001) surface upon annealing fits well with
the Wang~\textit{et al.}~\cite{wa-jpc-94-5966} interpretation of
the annealing of the low dosed PH$_{3}$/Si(001),  and contradicts
the interpretation of Kipp~\textit{et al.}~\cite{ki-prb-95-5843}
We therefore conclude, following Wang~\textit{et al.}, that the
processes that occur as a result of the 350$^{\circ}$C anneal are
as follows; the PH$_{x}$ (x=2,3) molecules firstly undergo
complete dissociation. Each P atom then undergoes a substitution
reaction with one atom of the Si surface dimer to form a Si-P
heterodimer in the plane of the surface, thereby ejecting the
displaced Si atom onto the surface. At 350$^{\circ}$C the
displaced Si atoms are mobile enough to diffuse short
distances~\cite{mo-prl-91-1998} and subsequently form 1-D dimer
chains on the surface. The H atoms liberated from PH$_{x}$
dissociation are highly mobile at 350$^{\circ}$C and pair-up to
form the energetically favorable monohydride
dimers.~\cite{sh-jpc-96-4961}

\subsection{The changing appearance of the Si-P heterodimer with coverage}

\begin{figure}
\begin{center}
\includegraphics[width=8cm]{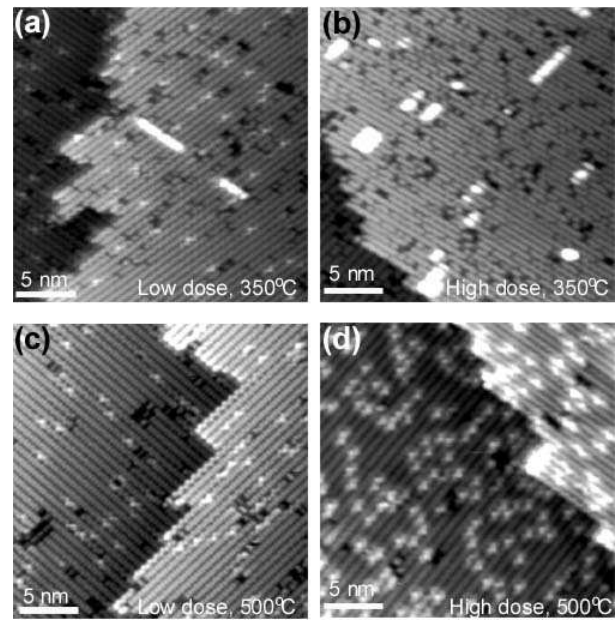}
  \caption{\label{DoseAndHeat}(a), (b) STM images of a Si(001) surface heated to 350$^{\circ}$C for 10 seconds, after a PH$_{3}$ dose of (a) 0.002 L and (b) 0.01 L.
  (c) and (d) show the same surfaces in (a) and (b) respectively, after a further 10 second anneal to 500$^{\circ}$C.
All images were obtained using a tunneling current of 0.13 nA, and
sample biases of -1.6 V except (d) which was -0.8 V.}
\end{center}
\end{figure}

We have found that the appearance of the Si-P heterodimer varies
greatly as a function of initial PH$_{3}$ coverage.
Figure~\ref{DoseAndHeat} shows a Si(001) surface after exposure to
2 different doses (0.002~L and 0.01~L) of PH$_{3}$ and subsequent
annealing of the dosed surfaces to 350$^{\circ}$C and then
500$^{\circ}$C. If we compare the two 350$^{\circ}$C anneals,
Fig.~\ref{DoseAndHeat}(a) and (b), we see that the surface that
was exposed to the higher PH$_{3}$ dose has significantly more Si
chains and the onset of 2D Si island formation
(Fig.~\ref{DoseAndHeat}(b)), as expected. After both the lower and
higher dosed surfaces have been subject to a further anneal of
500$^{\circ}$C the Si chains have now disappeared, see
Fig.~\ref{DoseAndHeat}(c) and (d). This can be explained since at
500$^{\circ}$C the diffusion of Si atoms on the surface is
sufficient that all the Si ad-dimers migrate to step edges to
which they preferentially bond.
We now consider what information we can determine about the Si-P
heterodimers, from Fig.~\ref{DoseAndHeat}. If we just consider
Fig.~\ref{DoseAndHeat}(c) and 3(d), which show the lower and
higher dosed surfaces after annealing to 500$^{\circ}$C, the data
fits our explanation of the evolution of the PH$_{3}$ dosed
Si(001) surface i.e. the more PH$_{x}$ (x=2,3) initially adsorbed
on the surface, the more Si-P heterodimers are observed on the
surface after the 500$^{\circ}$C anneal. However, the
350$^{\circ}$C anneals of the lower and higher dosed surfaces are
not so easy to understand. We repeatedly observe that lower dosed,
350$^{\circ}$C annealed surfaces such as that of
Fig.~\ref{DoseAndHeat}(a) show moderate densities of Si-P
heterodimers and monohydrides, as expected. Conversely, the higher
dosed, 350$^{\circ}$C annealed surfaces such as that of
Fig.~\ref{DoseAndHeat}(b) appear to show a high density of
monohydrides and no Si-P heterodimers. We know from the
500$^{\circ}$C anneal of the higher dosed surfaces that Si-P
heterodimers are abundant, so we can conclude that a high density
of monohydrides inhibits the observation of Si-P heterodimers with
the STM. It is known that monohydride dimers exhibit very little
buckling~\cite{ps-pc-private} and thus pin neighboring Si dimers
in an unbuckled geometry. Hence, if the density of monohydrides is
sufficiently high it is reasonable to assume that almost all the
surface Si dimers are pinned, even those in the vicinity of Si-P
heterodimers. As a consequence, the characteristic zig-zag
appearance of the Si-P heterodimer will be suppressed. It is now
interesting to look again at Fig.~\ref{350A}(d) which was obtained
by annealing a buried P layer (with no H present) to
600$^{\circ}$C. Whilst a sufficient number of P atoms have
diffused to the surface to yield a high density of Si-P
heterodimers, the surface is free of H. In this image the Si-P
heterodimers are clearly visible, which is consistent with the
observation that it is a high density of monohydrides on the
surface that inhibits the STM observation of Si-P heterodimers.

Our final anneal of the PH$_{3}$/Si(001) low dosed surface (not
shown) was performed at 700$^{\circ}$C and resulted in the
disappearance of the Si-P heterodimers.  By comparing our results
with that from the annealing of a saturation dosed
PH$_{3}$/Si(001) surface,~\cite{li-ss-99-7} we conclude that the P
atoms have desorbed from the surface, leaving only features
inherent to the clean Si(001) surface.
\subsection{Identification of the Si-P heterodimer using empty state imaging}

In the previous section we showed that for high coverages of
monohydrides, the identification of Si-P heterodimers in filled
state images was not possible. However, we will show that we can
clearly identify Si-P heterodimers, at any coverage, using empty
state imaging. Figure~\ref{dualbias} shows (a) filled- and (b)
empty-state images of the same area of a Si(001) sample after a
0.002 L PH$_{3}$ dose and a 500$^{\circ}$C anneal. We have chosen
a low PH$_{3}$ dose and an anneal temperature sufficiently high
that ejected Si dimer chains and monohydrides have been completely
removed from the surface, so that we can image Si-P heterodimers
in isolation from other features. Figure~\ref{dualbias}(b) shows
that in empty state images the Si-P heterodimer appears as a
bright circular protrusion on one side of a dimer row with a
slight depression on the other side of the row. The height of the
protrusion, measured with respect to the surface dimers, varies
linearly with bias between $0.50\pm0.05$~\AA~at +0.8~V to
$0.10\pm0.02$~\AA~at +2.0~V. Dimer buckling in empty state images
of Si(001) is not observed for sample biases greater than
+0.8~V.~\cite{qi-prb-99-7293}

With the Si-P heterodimer well characterized in both filled and
empty state imaging we can now also interpret images obtained
after higher PH$_{3}$ doses but a lower anneal temperature, where
H is still present. Figures~\ref{dualbias}(c) and (d) are dual
bias images of a Si(001) surface that has been dosed with 0.013 L
of PH$_{3}$ and annealed to 350$^{\circ}$C. The filled state image
shows a number of Si chains and monohydrides but no Si-P
heterodimers are observed. We see that the ejected Si chains have
a characteristic minima running lengthways along their center in
the empty state image, as
expected.~\cite{wa-jpc-94-5966,sw-ss-97-195} However, despite the
difficulty in resolving features due to the extreme brightness of
the ejected Si, in the empty state image we still see circular
protrusions characteristic of the Si-P heterodimers, as well as
the Si chains and the monohydrides. This ability to identify
incorporated P atoms regardless of the density of other features
on the surface is a useful tool for locating P atoms in Si,
towards the goal of realizing atomic-scale
devices.~\cite{tu-ijcta-00-553,sc-prl-03-136104}

\section{Conclusion}

We have used scanning tunneling microscopy (STM) and Auger
electron spectroscopy (AES) to study the incorporation of P into
Si(001) by dosing with small quantities of PH$_{3}$ and annealing.
The Si-P heterodimer appears as a zig-zag feature when imaged in
filled state STM images due to the static dimer buckling induced
by the heterodimer. The Si-P assignment was confirmed from the
direct correlation between the number of zig-zag features imaged
by STM and the fraction of P at the surface measured by AES. The
presence of moderate numbers of monohydrides on the surface
removes the buckling induced by the Si-P heterodimer and renders
the heterodimer invisible in filled state images. We characterized
the Si-P heterodimer using empty state imaging and show that, in
the empty state the Si-P heterodimers are visible even with a high
coverage of monohydrides. The demonstration of the incorporation
of individual P atoms into the Si(001) surface, and subsequent
characterization using STM, is an important step towards
fabrication of nanoscale electronic devices based on controlled
atomic-scale doping of Si.

\begin{acknowledgements}
This work was supported by the Australian Research Council, the
Australian Government and by the US National Security Agency
(NSA), Advanced Research and Development Activity (ARDA) and the
Army Research Office (ARO) under contract number DAAD19-01-1-0653.
LO acknowledges a Hewlett-Packard Fellowship.
\end{acknowledgements}

\end{document}